\documentclass[11pt]{article}
\usepackage[T1]{fontenc}
\usepackage[utf8]{inputenc}
\usepackage{lmodern}
\usepackage{microtype}
\usepackage[margin=1in]{geometry}
\usepackage{graphicx}
\usepackage{booktabs}
\usepackage{array}
\usepackage{natbib}
\usepackage[hidelinks]{hyperref}
\usepackage{authblk}
\usepackage{caption}
\usepackage{setspace}
\title{Evaluating AI Tutoring at the Speed of Innovation:\\Practitioner-Led Micro-Randomised Trials of an AI Tutoring Platform in GCSE Science}
\author[1]{Wayne Harrison}
\author[1]{Rahil Khowaja}
\author[1]{Emma Dobson}
\author[1,2]{Germaine Uwimpuhwe}
\author[1,2]{Steve Higgins\thanks{Corresponding author: \href{mailto:s.e.higgins@durham.ac.uk}{s.e.higgins@durham.ac.uk}}}
\affil[1]{What Worked Education}
\affil[2]{Durham University}
\date{}
\begin{document}
\maketitle

\begin{abstract}
Artificial intelligence (AI) systems in education are developing on timescales that sit uneasily with conventional evaluation. By the time a large-scale trial has been designed, delivered, analysed and published, the technology under study may have changed materially. This creates a temporal problem for evidence-informed education: the need for timely evidence can encourage reliance on weak observational or usage data, while conventional rigorous evaluation may produce evidence too slowly to guide rapidly evolving practice. We examine practitioner-led micro-randomised controlled trials (micro-RCTs) as one response to this problem. The empirical case is a four-week multisite individually randomised evaluation of Medly, an AI-powered tutoring platform, in GCSE Biology, Chemistry and Physics in English secondary schools. Of 929 students completing baseline assessment, 644 completed post-testing. In the primary intention-to-treat analysis, students allocated to Medly achieved higher post-test attainment than students undertaking business-as-usual self-directed revision (Hedges' g = 0.33, 95\% CI 0.18 to 0.48). Positive estimates were observed in Physics (g = 0.31), Chemistry (g = 0.32) and Biology (g = 0.52), with no evidence of differential impact by Pupil Premium status. Greater platform engagement was associated with higher attainment, but these post-randomisation analyses are treated as exploratory rather than causal. Attrition was substantial (30.7\%), the outcome measures were curriculum-aligned rather than standardised, and process evaluation response was limited. We therefore interpret the findings as preliminary. More broadly, we argue that the value of micro-RCTs for educational AI lies not in replacing definitive evaluation with small studies, but in enabling a rapid, cumulative evaluation architecture in which randomised estimates can be generated, replicated and updated as technologies and their implementation evolve.
\end{abstract}


\section{Introduction: the temporal problem of evaluating educational AI}

Artificial intelligence has intensified a long-standing problem in educational technology: innovation can move faster than the evidence used to judge it. The problem is not only evaluation lag (the interval between a technology becoming available and credible evidence becoming available about it) but also intervention drift. During that interval, the system itself may change: underlying models, interfaces, feedback routines, curriculum content, safeguards and patterns of implementation can all evolve. An internally valid estimate may therefore describe a technological configuration that is no longer the one encountered by schools.

This creates a temporal problem for evidence-informed education. Schools and developers need timely information about whether a new system is likely to improve learning and under what conditions. In the absence of timely causal evidence, decisions may be driven by uptake, satisfaction, engagement metrics, before-and-after comparisons or commercial claims. Yet accelerating evaluation by relaxing standards of causal inference creates the opposite problem: evidence may arrive quickly but provide a weak basis for attributing observed changes to the technology \citep{dawson2018}. For rapidly evolving AI, methodological credibility and temporal proximity both contribute to the practical relevance of evidence.

\subsection{The moving intervention}

Conventional evaluation works most comfortably when an intervention can be specified with sufficient stability to support a lengthy cycle of protocol development, recruitment, implementation, analysis and publication. Educational AI complicates this assumption. The intervention may be better understood as a versioned configuration of technological capabilities, pedagogical functions and implementation practices at a particular point in time: a moving intervention rather than a fixed product.

Technological change is only one source of movement. The realised intervention also changes through implementation. A platform intended as homework may be used during lessons; pupils may encounter it primarily through mobile devices rather than computers; teachers may vary in training and support; and learners may respond differently to scaffolded interaction. Evaluating AI therefore requires documentation not only of software version but of the pedagogical framing and implementation conditions instantiated in the version that is actually studied.

\subsection{Rapid randomisation as a cumulative architecture}

The methodological challenge is consequently not a simple choice between speed and rigour. It is to develop evaluation architectures that preserve credible counterfactual inference while operating on timescales closer to technological development. Practitioner-led micro-randomised controlled trials (micro-RCTs) offer one possible response. Their principal methodological advantages are not their size and speed, but their repeatability. Focused randomised comparisons can be embedded in routine practice, completed relatively quickly and repeated across teachers, schools, topics, cohorts and successive technological versions.

This suggests a progression from identifying signal to replication to accumulation. An initial micro-RCT provides a provisional causal signal. Subsequent trials test whether that signal replicates under changed populations, contexts, implementation conditions or product versions. Across cycles, estimates can be synthesized or aggregated cumulatively while retaining their individual variation. The aim is not to replace larger effectiveness trials with a collection of small studies, but to allow randomised evidence to enter the development cycle earlier and to become progressively more informative.

\subsection{AI tutoring and the present study}

AI applications in education include adaptive learning, automated assessment, intelligent tutoring and generative conversational support with a rapidly expanding empirical literature examining their relationship with student learning outcomes \citep{sasikala2024}. In science education, these systems may provide immediate feedback, scaffold multi-step reasoning and tailor practice to current understanding \citep{chng2023,cooper2023}. Reviews of intelligent tutoring systems generally report positive attainment effects, although estimates vary by educational phase, subject, comparison condition and design \citep{steenbergenhu2013,steenbergenhu2014,heeg2023,almasri2024,huang2025}. Recent experimental studies of generative-AI tutoring, including \citet{kestin2025} in undergraduate physics and \citet{learnlm2025} in UK secondary mathematics, further strengthen the case for direct outcome evaluation. Evidence remains comparatively sparse for curriculum-specific AI systems in school science under ordinary school conditions, though this is a developing field \citep{herdliska2024,sari2024}. To our knowledge, this is the first randomised controlled evaluation of a generative-AI tutoring system designed specifically to support students' constructed written responses to curriculum-aligned examination questions in Science.

The empirical example in this paper is an independent evaluation of Medly, an AI-powered tutoring platform providing personalised, curriculum-aligned practice and feedback. Parallel practitioner-led trials were conducted in GCSE (General Certificate of Secondary Education) Biology, Chemistry and Physics. The immediate questions were whether allocation to Medly improved attainment relative to business-as-usual self-directed revision, whether effects differed by disadvantage status, and how engagement related to outcomes. The wider question is what this form of rapid randomised evaluation can contribute to an evidence system capable of learning while the technology itself continues to develop and iterate.

\section{Methods}

\subsection{Design and setting}

The study was a multisite individually randomised controlled trial implemented as parallel practitioner-led micro-RCTs in mainstream secondary schools in England. Three subject-specific trials were conducted: Biology (bioenergetics, photosynthesis and cellular respiration), Chemistry (principles of electrolysis), and Physics (basics of atomic structure). Each student participated in one subject trial only. The intervention period lasted four weeks. The focal content had already been taught, and the study examined revision rather than replacement of normal classroom instruction.

The WhatWorked Teachers platform supported trial delivery by handling randomisation, data-entry, validation and automated class-level ANCOVA reporting. This infrastructure was intended to reduce the technical burden on practitioners while separating allocation and analysis procedures from teacher judgement.

\subsection{Participants and randomisation}

Eligible students were in Year 9 or Year 10. Teachers were eligible where they taught the relevant science subject in a mainstream English secondary school, could complete the pre- and post-test within the trial window, and were not simultaneously implementing another structured intervention with the same classes. Before randomisation, students completed a 30-minute baseline assessment. The platform then randomised students individually to Medly or control, independently of the delivering teacher and research team. Allocation lists were used to configure subject access so that control students could not access the relevant Medly content during the trial.

\subsection{Intervention and counterfactual}

Medly is an AI-powered tutoring platform using multiple layers of large language models linked to an exam-curriculum-specific knowledge base. It combines structured lessons and practice of extended written answers, adaptive identification of areas of weakness, real-time adjustment of difficulty, and marking against relevant examination-board criteria. Intervention students were assigned one predefined activity each week for four weeks, with an expected duration of approximately 30 minutes per activity.

Control students were allocated 30 minutes per week of self-directed revision on the same focal topic using resources of their choice other than Medly. This was an active, ecologically realistic comparator rather than a no-treatment condition. Process-evaluation responses indicated that business-as-usual revision could include other digital platforms as well as printed and school-provided materials; this heterogeneity is considered in interpreting the treatment contrast.

\subsection{Outcomes and additional measures}

The primary outcome was attainment on a subject-specific GCSE-aligned assessment. Each assessment comprised five questions with a maximum score of 35. Pre- and post-tests were designed to be equivalent in conceptual coverage, difficulty and cognitive demand but used different items to reduce practice effects. Items were not available within the intervention during the trial. The assessment was not a standardised instrument; responses were AI-marked and then manually checked and approved by the class teacher, who could amend marks before scores were entered into the WhatWorked Teachers platform.

Pupil Premium status was recorded as a binary covariate using free-school-meal eligibility as a proxy in the analysis. (The Pupil Premium is extra funding provided to state-funded primary and secondary schools in England to help disadvantaged pupils achieve their full potential.) Engagement was measured from platform logs as the number of Medly questions answered during the intervention period.

\subsection{Statistical analysis}

The primary confirmatory analysis followed the intention-to-treat principle. Post-test attainment was modelled as the outcome with treatment allocation and pre-test score as predictors, using a mixed-effects framework to account for clustering at school level. The combined analysis produced an overall treatment estimate across the three science subjects, reported as Hedges' g with a 95\% confidence interval. Subject-specific models were also estimated for Physics, Chemistry and Biology.

Differential impact by Pupil Premium status was examined using a treatment-by-status interaction. Engagement analyses were exploratory because engagement was observed after randomisation and only for intervention students. Associations between questions answered and post-test attainment, adjusted for pre-test attainment, are therefore reported as associational rather than causal. The original evaluation also reported threshold-based estimates among increasingly engaged intervention participants. In this paper these are not interpreted as complier average causal effects, because conditioning on post-randomisation engagement does not by itself identify a causal complier effect.

\subsection{Process evaluation and ethics}

Teachers were invited to complete a process-evaluation survey covering implementation, practical barriers, confidence in the platform, patterns of use and business-as-usual revision. Six of 39 teacher trials returned a process survey, so these findings are used as indicative implementation evidence rather than as a representative account.

Student records in the evaluation platform were anonymised. The evaluation used a privacy-preserving linkage procedure for usage and disadvantage-status analyses. Activities were aligned with routine teaching and learning, and the protocol underwent internal ethics review and independent academic review. Active parental consent procedures can themselves introduce systematic bias in school-based research particularly where participation varies according to pupil characteristics \citep{shaw2015,liu2017}. Processing was undertaken under the UK GDPR public-interest basis and headteacher consent was used for participation in these low-risk routine educational activities.

\section{Results}

\subsection{Sample and attrition}

A total of 929 students completed the pre-test and 644 completed the post-test, giving overall attrition of 30.7\%. Post-test samples comprised 312 students in Physics, 217 in Chemistry and 115 in Biology. Among post-test completers, 332 were allocated to intervention and 312 to control. Attrition was higher in the control group (33.2\%) than in the intervention group (27.7\%), creating a potential source of bias that is considered in the limitations.

\begin{table}[htbp]
\centering
\caption{Participant flow by subject.}
\label{tab:flow}
\begin{tabular}{lrrr}
\toprule
Subject & Pre-test $n$ & Post-test $n$ & Loss to follow-up \\
\midrule
Physics & 505 & 312 & 38.22\% \\
Chemistry & 275 & 217 & 21.09\% \\
Biology & 149 & 115 & 22.82\% \\
Total & 929 & 644 & 30.68\% \\
\bottomrule
\end{tabular}
\end{table}

\subsection{Primary intention-to-treat effect}

In the combined primary analysis, allocation to Medly was associated with higher post-test attainment than allocation to business-as-usual revision. The adjusted mean post-test score was 13.95 in the intervention group and 11.39 in the control group. The adjusted coefficient was 2.56 marks (95\% CI 1.39 to 3.73), corresponding to Hedges' g = 0.33 (95\% CI 0.18 to 0.48).

\subsection{Subject-specific effects}

Positive effects were observed in all three subject-specific analyses. Physics produced g = 0.31 (95\% CI 0.11 to 0.51), Chemistry g = 0.32 (95\% CI 0.06 to 0.58), and Biology g = 0.52 (95\% CI 0.19 to 0.84). The subject-by-treatment interaction did not provide evidence that the treatment effect differed across subjects. The larger point estimate in Biology should therefore be treated as descriptive rather than as evidence of a reliably larger subject effect.

\begin{table}[htbp]
\centering
\caption{Subject-specific intention-to-treat estimates.}
\label{tab:subject}
\small
\begin{tabular}{lrrcc}
\toprule
Subject & Intervention $n$ & Control $n$ & Adjusted difference (95\% CI) & Hedges' $g$ (95\% CI) \\
\midrule
Physics & 164 & 148 & 2.93 (1.03, 4.83) & 0.31 (0.11, 0.51) \\
Biology & 59 & 56 & 2.93 (1.14, 4.71) & 0.52 (0.19, 0.84) \\
Chemistry & 109 & 108 & 1.95 (0.38, 3.52) & 0.32 (0.06, 0.58) \\
\bottomrule
\end{tabular}
\end{table}
\begin{figure}[htbp]
\centering
\includegraphics[width=\textwidth]{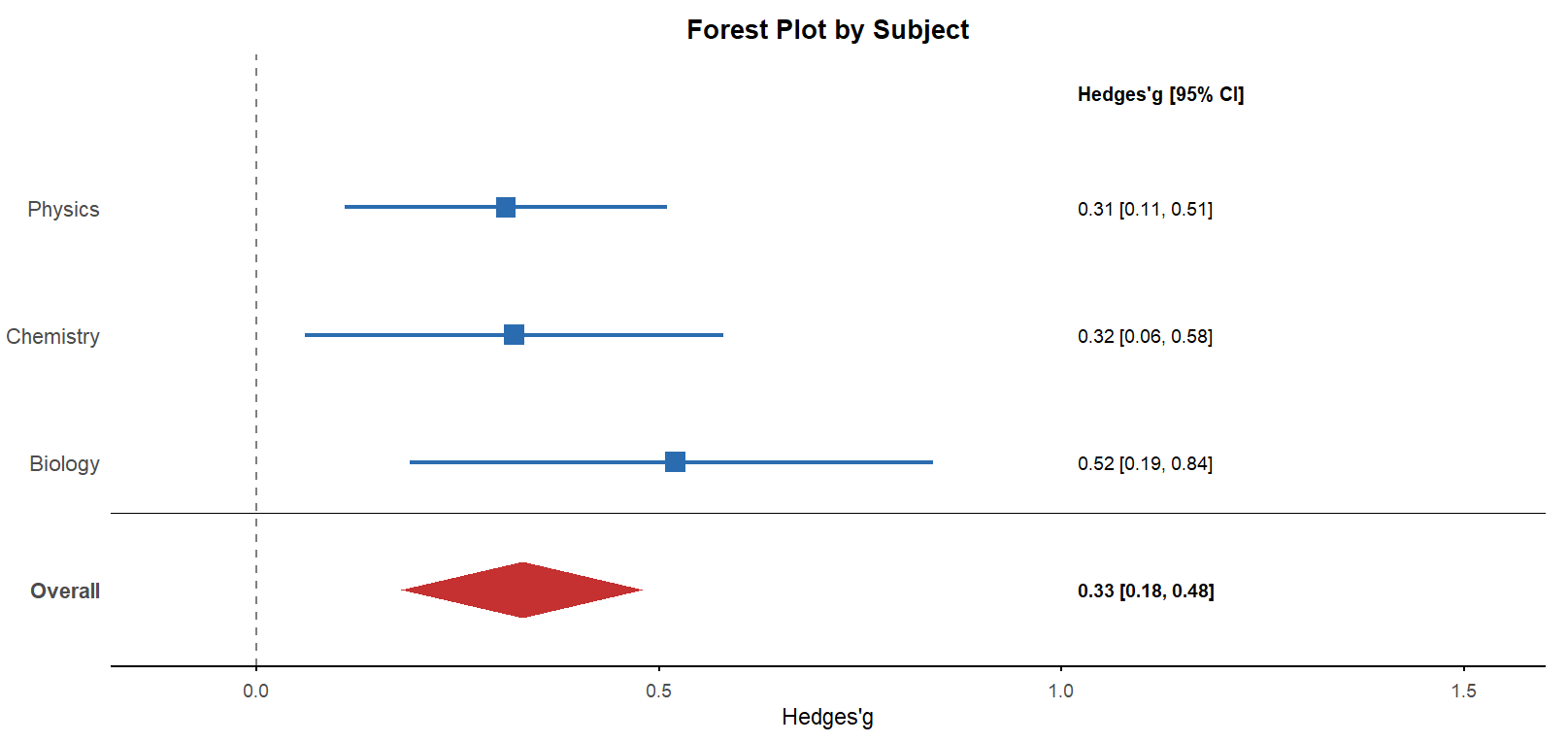}
\caption{Forest plot of overall and subject-specific Hedges' $g$ estimates with 95\% confidence intervals.}
\label{fig:forest}
\end{figure}

\subsection{Pupil Premium subgroup analysis}

There was no clear evidence that the effect of Medly differed by Pupil Premium status. The treatment-by-status interaction estimate was 0.57 marks (95\% CI -2.25 to 3.39). Exploratory stratified estimates were positive for both groups: g = 0.28 (95\% CI -0.04 to 0.59) among Pupil Premium students and g = 0.35 (95\% CI 0.18 to 0.52) among non-Pupil Premium students. The wider interval for the Pupil Premium subgroup is consistent with its smaller sample and should not be interpreted as evidence of absence of benefit.

\subsection{Engagement}

Within the intervention group, greater engagement was positively associated with post-test attainment after adjustment for baseline attainment. Each additional question answered was associated with approximately 0.18 additional post-test marks (95\% CI 0.13 to 0.22). Positive associations were also observed separately in Physics, Biology and Chemistry. Because engagement is a post-randomisation behaviour that may reflect motivation, prior attainment, access and other characteristics, these analyses cannot establish that increasing the number of questions answered would itself cause the observed increase in attainment.

The original evaluation additionally reported progressively larger treatment estimates after restricting the intervention group to students crossing thresholds of 10, 20 and 30 questions. This pattern is consistent with the hypothesis that effective use matters, but selection into higher engagement makes causal interpretation unsafe without an identification strategy for compliance. We therefore treat the threshold pattern as hypothesis-generating evidence for future trials.

\subsection{Implementation findings}

Only six of the 39 teacher trials returned process-evaluation surveys. Within this small sample, implementation varied substantially. Two teachers reported setting Medly as directed homework, while four used it during lesson time. Four had attended a Medly-led training session and two had not. This variation suggests that schools adapted the intended delivery model to local routines and access constraints.

The clearest practical issue was technical friction, especially mobile-device access and login. At the same time, respondents described responsive support from the Medly team. Teacher comments also suggested variation in pupils' willingness to work through scaffolded reasoning rather than seek immediate answers. These observations are particularly relevant to iterative AI evaluation: they identify implementation features that can be changed and retested in subsequent cycles.

Business-as-usual revision was heterogeneous. The six respondents described resources including Tassomai, Sparx Science, BBC Bitesize and printed revision materials. In at least some classes, therefore, the treatment contrast was not AI versus no digital support but Medly versus an already technology-rich revision environment.

\section{Discussion}

\subsection{A provisional causal signal}

The primary randomised estimate provides preliminary evidence that access to Medly improved short-term attainment on curriculum-aligned GCSE science questions relative to business-as-usual revision. The overall effect of approximately one third of a standard deviation is educationally meaningful if replicated, and the direction of effect was positive across Physics, Chemistry and Biology. There was no clear evidence of differential impact by Pupil Premium status.

The appropriate interpretation is nevertheless a provisional causal signal rather than a definitive effectiveness estimate. Attrition was high, the assessment was curriculum-aligned rather than independently standardised, follow-up was short, and process-evaluation data were available from only a small minority of participating teachers. Within a rapid cumulative model, these limitations do not make the initial estimate uninformative; they define what the next evaluation cycle needs to test more securely.

\subsection{The moving intervention and temporal relevance}

The broader methodological issue is that delay in AI evaluation can become more than a logistical inconvenience. If the model, interface, feedback logic or implementation arrangements change materially between trial launch and publication, the identity of the evaluated intervention becomes less stable. Evidence can retain high internal validity for the version studied while becoming progressively less informative about the version currently in use. Temporal proximity is therefore relevant to the warrant for applying an estimate to a rapidly evolving system.

This does not imply that every software update creates an entirely new intervention. A useful distinction is between technological implementation, instructional features and pedagogical intention. Interfaces and underlying models may change quickly; functions such as adaptive practice, diagnostic feedback and scaffolding may be more stable; and the broader pedagogical intention, for example, supporting pupils to retrieve knowledge, identify errors, reason through problems and improve responses, may be more stable still. Cumulative evaluation requires enough version information to identify technological change while also documenting the pedagogical conditions that permit learning across versions.

\subsection{Repeatability rather than size and cost}

This perspective clarifies the role of micro-RCTs \citep{harrison2025}. Their value for educational AI is not primarily that they are small or inexpensive. It is that repeated causal estimation becomes feasible. Random allocation is particularly important in fast-moving technology markets because early adopters, enthusiastic teachers and highly engaged pupils are unlikely to be representative. A realistic counterfactual is equally important. In the present study, control students undertook ordinary self-directed revision, and some used established digital platforms; the estimate therefore concerns added value over existing revision practice rather than comparison with inactivity.

Rapid evaluation also requires disciplined limits on interpretation. Platform analytics are immediate and granular, but analysis of post-randomisation usage does not provide causal inference. The positive association between Medly engagement and attainment is compatible with a treatment mechanism, but it may also reflect motivation, access, prior capability or other characteristics that influence engagement. The randomised intention-to-treat contrast should therefore remain the causal centre of the evidence, with engagement data used to generate hypotheses for subsequent tests.

\subsection{Signal, replication and accumulation}

A rapid cumulative architecture can be understood as three linked stages. The first study generates a signal. Replication then asks whether the direction and magnitude of effect recur in another cohort, setting, implementation period or technological version. Accumulation asks a broader question: what distribution of effects emerges across these trials, and what features of version, pedagogy, population and implementation help to explain the variation?

This shifts the mature evidence question from 'What is the effect of Medly?' towards 'What distribution of effects is produced when this evolving AI-supported pedagogical approach is implemented across pupils, teachers, contexts and technological versions?' Successive micro-RCTs can be represented initially through forest plots and cumulative meta-analysis, retaining individual estimates rather than allowing a pooled mean to obscure heterogeneity. More formal hierarchical or sequential models could subsequently represent version and implementation explicitly.

The same logic gives implementation evidence a more active role. Technical friction, device access, login processes and the way teachers position an AI tutor affect whether its intended pedagogical functions are achieved. Rapid cycles allow such findings to become hypotheses for the next randomised iteration: modify the implementation, document the new configuration, repeat the trial and examine whether engagement and attainment change.

\subsection{Implications for an adaptive evidence base}

For developers, evaluation can become part of responsible product development rather than a certification exercise conducted after development. For schools, practitioner-led randomisation offers a way to contribute to a shared evidence base while testing questions in authentic settings. For evaluators, the challenge is to combine rapidity with cumulative inference through common protocols, secure randomisation, transparent reporting, prospective specification where feasible, consistent core outcomes and explicit version documentation.

The methodological challenge posed by educational AI is therefore not simply how to evaluate new technologies more quickly. It is how to construct an evidence system capable of learning while the intervention itself is learning. Individual micro-RCTs provide provisional estimates; replication tests their stability; cumulative synthesis reveals their distribution; and implementation evidence helps explain why effects change. Technological iteration need not render previous evidence obsolete if successive versions, pedagogical functions and implementation conditions are represented explicitly in the accumulating evidence base. For rapidly evolving educational AI, the evidence base may need to become adaptive too.

\section{Limitations}

First, 30.7\% of baseline participants did not complete the post-test, and attrition differed modestly between treatment arms. Although the intention-to-treat analysis preserves original allocation among observed outcomes, missing post-test data may bias the estimate if attrition is related jointly to allocation and potential outcomes. Replication should prioritise reducing attrition and should pre-specify sensitivity analyses for missing outcomes.

Second, the primary assessments were designed around GCSE content but were not standardised external measures. AI-generated marking was checked by teachers, which provides a practical safeguard but does not substitute for independent blinded outcome assessment. Future trials would be strengthened by independent marking or moderation and, where feasible, externally validated or examination-linked outcomes in the longer term.

Third, the four-week intervention estimates short-term topic learning rather than persistence, transfer or examination performance. Longer follow-up is needed.

Fourth, only six of 39 teacher trials contributed process-evaluation data. Implementation findings are therefore illustrative and should not be generalised to all participating classes.

Fifth, the engagement analyses are vulnerable to post-randomisation selection. They identify a robust association but not a causal dose-response function. A future study could randomise encouragement, assigned practice intensity or another valid instrument for engagement if a causal estimate of compliance is required.

Finally, the study evaluates a particular configuration of Medly at a particular point in its development. This is precisely the problem addressed by the paper's wider argument: evidence for rapidly evolving AI should be version-aware and updated through replication.

\section{Conclusion}

This study provides preliminary randomised evidence that an AI-powered tutoring platform can improve short-term attainment in extended written answers in GCSE science relative to business-as-usual revision. The overall estimate was positive, with positive subject-specific estimates in Physics, Chemistry and Biology. Given attrition, outcome measurement and short follow-up, the result warrants replication rather than a definitive claim of effectiveness.

The wider contribution is an evaluation model suited to a moving intervention. Practitioner-led micro-RCTs can generate an early causal signal; repeated trials can test whether that signal replicates as populations, implementation and technology change; and cumulative synthesis can reveal both the average effect and its distribution. The aim is not evaluation at the speed of innovation at the expense of rigour, but rather an evidence system able to learn at a pace closer to the technologies it evaluates.

\section*{Declarations}

Funding: The evaluation was funded by Medly who commissioned WhatWorked Education to undertake the evaluation. Medly provided access to the schools who were using their platform and usage data. The participating teachers entered the pre- and post-test data.

Acknowledgements: The WhatWorked team would like to thank the staff at Medly for their support in working collaboratively to ensure the success of the evaluation and for the opportunity to publish the results and to thank the collaborating teachers who used the micro-RCT platform.

Competing interests: There were no competing interests or connections between Medly and the WhatWorked team other than through the evaluation.

Author contributions: Wayne Harrison: Investigation, Conceptualisation, Project administration, Methodology, Writing -- original draft \& review \& editing. Rahil Khowaja: Methodology, Data Curation, Formal analysis. Emma Dobson: Investigation, Methodology, Writing -- review \& editing. Germaine Uwimpuhwe: Methodology, Formal analysis, Validation, Writing -- review \& editing. Steve Higgins: Conceptualization, Methodology, Supervision, Writing -- review \& editing.

Data availability: For enquiries about the data and code, please contact the corresponding author: s.e.higgins@durham.ac.uk.

Ethics: The micro-RCTs underwent internal ethics review at design stage and independent academic review before launch; student data were anonymised and processed under the UK GDPR public-interest basis.

\end{document}